\documentclass[manuscript,screen,nonacm]{acmart}

\usepackage{graphicx} 
\usepackage{listings}
\usepackage[edges]{forest}
\usetikzlibrary{arrows.meta}

\colorlet{punct}{red!60!black}
\definecolor{background}{HTML}{EEEEEE}
\definecolor{delim}{RGB}{20,105,176}
\colorlet{numb}{magenta!60!black}

\lstdefinelanguage{json}{
    basicstyle=\normalfont\ttfamily,
    numberstyle=\scriptsize,
    stepnumber=1,
    numbersep=8pt,
    showstringspaces=false,
    breaklines=true,
    frame=lines,
    backgroundcolor=\color{background},
    literate=
     *{0}{{{\color{numb}0}}}{1}
      {1}{{{\color{numb}1}}}{1}
      {2}{{{\color{numb}2}}}{1}
      {3}{{{\color{numb}3}}}{1}
      {4}{{{\color{numb}4}}}{1}
      {5}{{{\color{numb}5}}}{1}
      {6}{{{\color{numb}6}}}{1}
      {7}{{{\color{numb}7}}}{1}
      {8}{{{\color{numb}8}}}{1}
      {9}{{{\color{numb}9}}}{1}
      {:}{{{\color{punct}{:}}}}{1}
      {,}{{{\color{punct}{,}}}}{1}
      {\{}{{{\color{delim}{\{}}}}{1}
      {\}}{{{\color{delim}{\}}}}}{1}
      {[}{{{\color{delim}{[}}}}{1}
      {]}{{{\color{delim}{]}}}}{1},
}

\AtBeginDocument{%
  \providecommand\BibTeX{{%
    \normalfont B\kern-0.5em{\scshape i\kern-0.25em b}\kern-0.8em\TeX}}}

\setcopyright{acmcopyright}
\copyrightyear{2023}
\acmYear{2023}
\acmDOI{XXXXXXX.XXXXXXX}

\acmPrice{}

\usepackage{xpatch}

\makeatletter
\xpatchcmd{\ps@firstpagestyle}{Manuscript submitted to ACM}{}{\typeout{First patch succeeded}}{\typeout{first patch failed}}
\xpatchcmd{\ps@standardpagestyle}{Manuscript submitted to ACM}{}{\typeout{Second patch succeeded}}{\typeout{Second patch failed}}    \@ACM@manuscriptfalse
\makeatother

\renewcommand\footnotetextcopyrightpermission[1]{} 
\setcopyright{none}
\begin{document}

\title{SSI4IoT: Unlocking the Potential of IoT Tailored Self-Sovereign Identity}

\author{Thusitha Dayaratne}
\email{thusitha.dayaratne@monash.edu}
\orcid{0000-0003-0624-1967}
\affiliation{%
  \institution{Monash University}
  \city{Clayton}
  \state{Victoria}
  \country{Australia}
}

\author{Xinxin Fan}
\orcid{0000-0002-4592-6603}
\affiliation{%
  \institution{IoTeX}
  \city{Melno Park}
  \state{CA}
  \country{USA}
}

\author{Yuhong Liu}
\affiliation{%
  \institution{Santa Clara University}
  \city{Santa Clara}
  \state{CA}
  \country{USA}
}

\author{Carsten Rudolph}
\affiliation{%
  \institution{Monash University}
  \city{Clayton}
  \state{Victoria}
  \country{Australia}
}

\renewcommand{\shortauthors}{T. Dayaratne, X. Fan, Y. Liu and C. Rudolph}

\begin{abstract}
The emerging Self-Sovereign Identity (SSI) techniques, such as Decentralized Identifiers (DIDs) and Verifiable Credentials (VCs), move control of digital identity from conventional identity providers to individuals and lay down the foundation for people, organizations, and things establishing rich digital relationship. The existing applications of SSI mainly focus on creating person-to-person and person-to-service relationships, whereas person-to-device and device-to-device interactions have been largely overlooked. In this paper, we close this gap by identifying a number of key challenges of applying SSI to the Internet of Things (IoT) and providing a comprehensive taxonomy and usage of VCs in the IoT context with respect to their validity period, trust and interoperability level, and scope of usage. The life-cycle management of VCs as well as various optimization techniques for realizing SSI in IoT environments are also addressed in great detail. This work is a noteworthy step towards massive adoption of SSI for securing existing and future IoT applications in practice. 
\end{abstract}




\maketitle

\section{Introduction}
The Internet of Things (IoT) is considered one of the most rapidly growing technologies in recent years. The number of IoT devices is forecasted to reach 16.3 billion by the end of 2023, marking an increase of 16\% over the count in 2022 \cite{IoTAnalytics}. IoT devices are equipped with sensors and firmware/software that manage their functions, computation, and communication. Advancements in IoT devices have enabled more efficient, effective, and automated processes, transforming traditional human-centric manual approaches in various sectors such as energy grids, transportation, agriculture, healthcare, and supply chain management. 

IoT manufacturers commonly rely on the centralized Public-Key Infrastructure (PKI) as well as hard-coded keys and digital certificates \cite{9225520,hoglund2020pki4iot} to bring trust to their IoT applications. However, the common concerns such as single point of failure, limited control over identities, privacy issues, scalability, and complexity make PKI-based approaches less favorable in the IoT context. Furthermore, recent NIST's requirements for achieving a secure and reliable onboarding process \cite{fagan2023trusted}, where (1) devices need to receive unique credentials at the time of deployment instead of depending on pre-configured keys/certificates and (2) onboarding credentials need to be communicated through an encrypted channel, are difficult to achieve with traditional PKI-based approach.

In recent years, self-sovereign identity (SSI) has gained significant attention, providing entities with complete control over their digital identity. The W3C's endorsement of Decentralized Identifiers (DID) \cite{didw3c} and Verifiable Credentials (VC) \cite{vcw3c} has further accelerated adoption of SSI, providing a decentralized and trustworthy solution without relying on a central authority. Many works have already adopted the SSI concept to provide reliable identity solutions for humans and devices. Nevertheless, not all devices possess sufficient capabilities and resources to handle the SSI functionalities. In particular, most of the IoT devices have limited computation and storage capabilities compared to common devices such as laptops, mobile phones, and computers~\cite{ren2017serving,pierleoni2019amazon}. Thus, despite the effectiveness of the SSI concept in terms of DID and VC, the use of these concepts in the IoT context has raised potential challenges. In particular, some argue that there is no alternative decentralised solution, which can replace the traditional certificate-based public key infrastructure that is suitable for use in IoT \cite{mahalle2020rethinking}.

Acknowledging the significant gaps that thwart the realization of practical SSI solutions in the IoT domain, in this paper, we analyze approaches towards the realization of an IoT friendly SSI paradigm. In particular, we present a design matrix that can be leveraged in designing VCs for IoT use cases, explorer potential VC issuers and common VCs that overlap in diverse IoT scenarios.
Further, we analyze the aspects of the life-cycle management of VCs and provide insights into diverse optimization techniques essential for the effective implementation of SSI in IoT environments. This contribution is expected to pave the way for researchers and industry practitioners to unlock the full potential of SSI in securing IoT applications. 


The rest of the paper is organized as follows. Section~\ref{sec:preliminaries} gives an overview of self-sovereign identity, decentralized identifier, and verifiable credential. In Section~\ref{sec:related}, we present the related work. Section~\ref{sec:vciot} provides a comprehensive taxonomy and usage of VCs in the IoT context based on three primary categories of factors, followed by the discussion of VC life-cycle management in Section~\ref{sec:lifecycle}. Section~\ref{sec:optimization} addresses potential optimization techniques when implementing SSI in IoT environments. In Section~\ref{sec:discussion}, we discuss a variety of challenges and concerns when applying SSI to the IoT. Finally, we conclude this paper in Section~\ref{sec:conclusion}.

\section{Preliminaries}
\label{sec:preliminaries}

\subsection{Self-Sovereign Identity (SSI}

\begin{figure}
    \centering
    \includegraphics[width=0.8\linewidth]{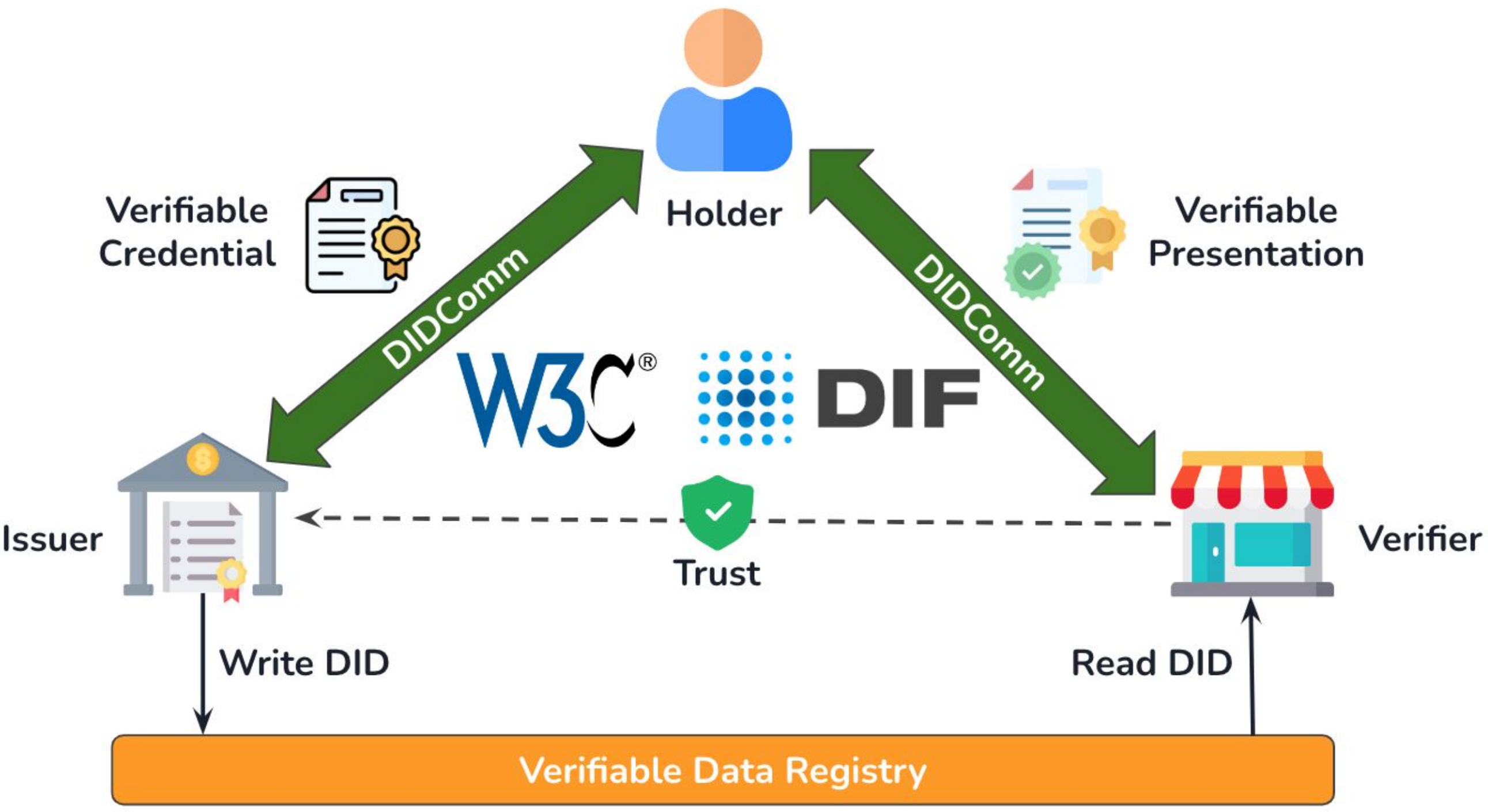}
    \caption{Three pillars of the SSI model}
    \label{fig:ssi_model}
\end{figure}

Digital identity management has undergone significant evolution since its inception to ensure the security, control, and portability of identities. The most prevalent model, the centralized or isolated identity model, involves a single entity owning and controlling all identities, restricting their validity to a specific domain~\cite{Tobin2017, Ferdous2019}. To address the portability limitations of this model, the federated identity model was introduced, allowing the reuse of a single identity across various domains \cite{Tobin2017, gaedke2005modeling}. Nonetheless, users in both models lack control over their identities. The advent of SSI represents the latest evolution of the identity journey. SSI empowers individuals with complete control over their identities, ensuring both absolute portability and enhanced security. In particular, the SSI model remains resilient in the face of centralized control and single point of failure, distinguishing it from previous identity models \cite{toth2019self, Tobin2017}. The main pillars of the SSI concept DIDs, VCs, and verifiable data registry are shown in Figure \ref{fig:ssi_model}. While blockchain technology often complements the SSI model, the SSI model itself is agnostic to blockchain technology.

\subsection{Decentralized Identifier (DID)}
A DID is a unique and persistent global identifier. DIDs enable verifiable and decentralized digital identities for entities \cite{didw3c} that are under total control of the identity owner, instead of relying on a central authority. The W3C approved the DID specification as a recommendation in late June 2022. The DID representation conforms with the Universal Resource Identifier (URI) scheme with three mandatory components as below.
\[
\color{blue}did:\color{magenta}<did\textnormal{-}method>:\color{teal}<method\textnormal{-}specific\textnormal{-}identifier>
\]

A DID resolves to a DID document, which is a JavaScript Object Notation-Linked Data (JSON-LD) object. The DID document possesses public keys for cryptographic operations, including authentication and verification. The DID document can also contain service endpoints, which can be used to advertise different services provided by the entity or the controller of the entity. Listing \ref{lst:diddoc} depicts a sample DID document.

\begin{lstlisting}[caption = Structure and content of a sample DID document,captionpos=b,label=lst:diddoc,language=json]
{
    "@context": [
        "https://www.w3.org/ns/did/v1",
        "https://w3id.org/security/suites/ed25519-2018/v1"
    ],
    "id": "did:sov:123456789",
    ...
    "verificationMethod": [{
        "type": "Ed25519VerificationKey2018",
        "id": "did:sov:123456789#key-1",
        "publicKeyBase58": "......"
    }],
    "authentication": [
        "did:sov:123456789#key-1"
    ],
    "service": [{
        "id": "did:sov:123456789#mqtt",
        "type": "MqttEndpoint",
        "serviceEndpoint": "broker.example.mqtt.com"
    },
    {
        "id": "did:sov:123456789#http",
        "type": "HttpEndpoint",
        "serviceEndpoint": "example.com"
    }]
}
\end{lstlisting}

\begin{figure}[H]
    \centering
    \includegraphics[width=0.8\linewidth]{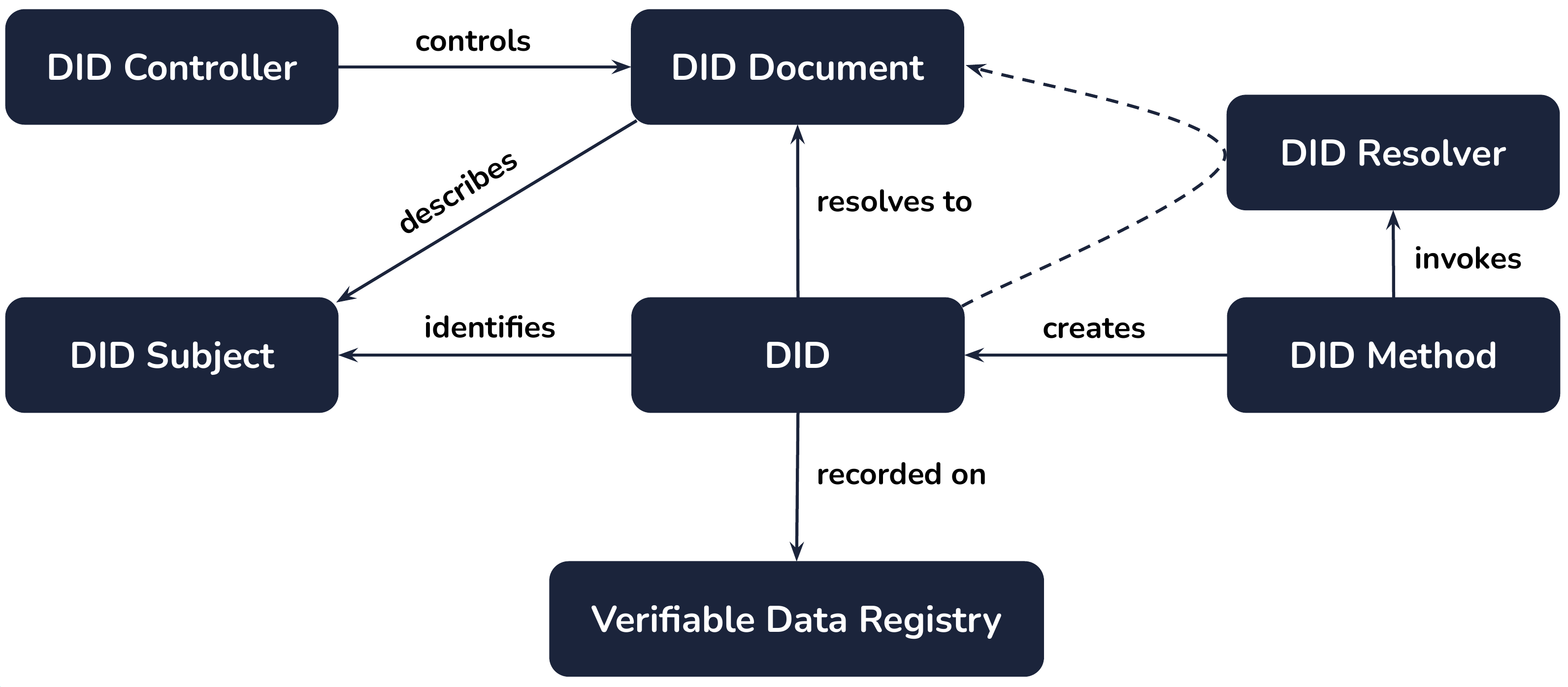}
    \caption{High level view of the DID ecosystem}
    \label{fig:did_architecture}
\end{figure}

Figure \ref{fig:did_architecture} provides an overview of the Decentralized Identifier (DID) architecture. A DID can reference any entity, including both physical and logical entities. The associated keys and endpoints of the referenced entity are detailed in the DID document. These documents are stored in publicly accessible, verifiable data registries, such as blockchains or distributed file systems like the InterPlanetary File System (IPFS). In most cases, DID documents should be publicly accessible and can be retrieved via the DID resolver. This enables communication initiation between stakeholders and the entity, similar to digital certificates in a PKI system. The DID itself is assumed to be immutable after its creation. However, the keys and endpoints linked to the DID can be updated within the DID document. By default, the entity is the owner/controller of the DID document and has full control over its updates. The $controller$ property can override this default and specify separate controllers for the DID document. For example, in the context of the IoT, a device or sensor might lack the capability to manage its own DID document. In such cases, the device or sensor owner assumes control over the device's DID document.

\subsection{Verifiable Credential (VC)}
A VC represents statements made by an issuer in a tamper-evident and privacy-respecting manner \cite{vcw3c}. Specifically, VCs provide portable and provable claims about an entity that can be cryptographically verified. The exchange of VCs builds trust among DID-identified peers. The W3C maintains the VC specifications, which consist of three components: Metadata, claims, and proof as shown in Listing \ref{lst:vc_sample}. Metadata describes meta-information about the VC, such as the issuer, expiration date, and public keys for verification. It also includes a set of claims about the entity and proofs. 

\begin{lstlisting}[caption = Structure of a simple VC,captionpos=b,label=lst:vc_sample,language=json]
{
  "@context": [
    "https://www.w3.org/2018/credentials/v1",
    ..
  ],
  "id": "did:iot:vc:123456789",
  "type": ["VerifiableCredential", "IoTVC"],
  "issuer": "did:iot:manufacturer:123456789",
  "validFrom": "2023-08-01T10:11:12Z",
  "credentialSubject": {
    "owner": "did:iot:user:123456789",
    "serialNo" : "1234567899",
    ..
  },
  ..
  "credentialSchema": { ..},
  "credentialStatus": { ..},
  "proof": { ..}
}
\end{lstlisting}

\begin{figure}[!htp]
    \centering
    \includegraphics[width=0.8\linewidth]{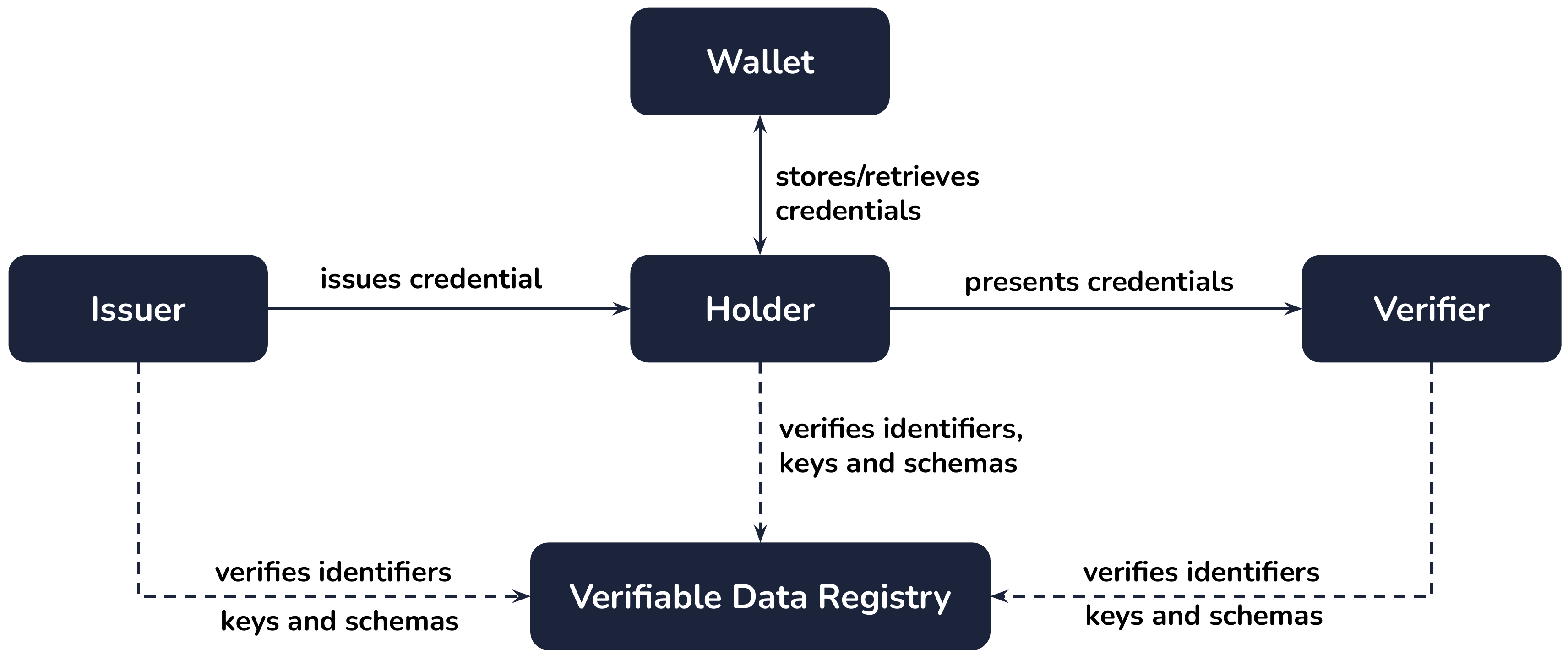}
    \caption{High level architecture of the verifiable credential ecosystem}
    \label{fig:vc_architecture}
\end{figure}

The VC ecosystem is depicted in Figure \ref{fig:vc_architecture}. It consists of three main entities (issuer, holder, and verifier) and a supporting verifiable data registry. Any entity can be a VC issuer. However, the trustworthiness of a particular VC depends on the issuer. Thus, VCs are usually issued by reputed or trusted entities in practical contexts. The holders store their VCs in a digital wallet and present them to the verifier upon request. Verifiers can verify the presented VCs without contacting the issuer, as verification keys can be accessed via the verifiable data registry. This prevents issuers from correlating the holder and the services that they access.

\subsubsection{Representations of VC}
While the W3C VC specification provides the semantic guidance with which all VCs must comply, the specification does not provide information on how to represent/render a VC. Thus, different vendors choose different methods to represent VCs. This section describes the commonly leveraged VC representations. 

JSON Web Token (JWT) is an open standard that utilizes JSON and base64 encoding to securely transmit information between different entities. Properties such as self-containment and compactness have made JWT the default choice for authentication and authorisation in web applications and APIs. In particular, well-established authentication and authorization frameworks, such as OAuth and OpenID Connect, are based on JWT. JWT based VC embeds the VC (without any proofs) in the JWT payload section and the VC proof as the JWT signature. These VC implementations use JWT to represent VC due to its familiarity among developers and the support of existing frameworks, standards, and libraries. However, while JWT has advantages, it lacks the required semantic interoperability needed to fully utilise the potential of VCs. Additionally, implementing advanced privacy features like zero-knowledge proofs (ZKPs) and predicates, which contribute to VC's superiority in privacy, can be complex with standard JWT based VC. However, there are initiatives such as SD-JWT-VC, which try to introduce selective disclosure on JWTs.

JSON-LD based VC representations address the limitations of JWT-based VCs. JSON-LD utilizes semantics for embedded data, enabling interoperability, portability, and comprehensibility across various contexts. It also facilitates the reuse of credential schemes, eliminating the need for different schemes to represent identical information. JSON-LD leverage LD signatures, allowing for a compact proof representation. Unlike JWT, JSON-LD signatures can cover individual attributes, enabling the combination of multiple VCs into a single verifiable presentation and supporting ZKPs. These attributes make JSON-LD a robust technology to implement VCs. Nevertheless, the expressiveness of JSON-LD requires extensive parsing using libraries that are capable of namespace and semantic resolution. 

Hyperledger defines AnonCreds (anonymous credentials) as a type of verifiable credential that supports privacy preservation capabilities through the use of ZKP. Unlike the other VC types, the production scale usage of AnonCreds around the world has made AnonCreds the default standard for ZKP-based verifiable credentials \cite{anoncredspec}. AnonCreds SDK is well matured and supports many popular languages, including Java, Rust, and Python.

\section{Related Work}
\label{sec:related}


Although the concept of VC is relatively new, researchers have already explored its use in various use cases. These works can be categorised as either involving human or device identity. However, relatively less work has analysed the use of VC and SSI concepts in a device context. More specifically, little work has explored the use of SSI in resource-constrained environments, such as IoT devices. In \cite{hosseini2023blockchain}, the authors provided an overview of existing identity management frameworks and their limitations. In particular, for resource-constrained IoT devices, the Authentication and Authorization for Constrained Environment (ACE) Working Group has proposed the ACE-OAuth framework based on OAuth2.0. However, this framework is dependent on central or federal service providers for identity management, access permissions, and data storage, which means that some third parties have control over all user data that can leak or change at will, and are vulnerable to hacks. To overcome this issue, the OAuth Working Group has introduced a decentralized service architecture, Decentralized OAuth, which relies on a peer-to-peer service architecture to achieve decentralization and smart contracts as a legal mechanism to bind services. Although this service architecture allows users to control their own data, there is no real-world implementation yet. In \cite{bartolomeu2019self}, the authors reviewed key use cases, technologies, and high-level challenges that SSI faces in the context of industrial IoT applications from technical, standardization, and organizational perspectives. The authors of \cite{fedrecheski2020self} have conducted a comprehensive comparison among PGP, X.509, and SSI for IoT environments, with detailed discussions of their data models for digital identity, key distribution, and attributes. With its numerous benefits, the adoption of SSI in resource-constrained IoT environments also faces several key challenges, including high computation, communication and storage overhead, device traceability, and availability. Similarly, in \cite{kortesniemi2019improving}, the authors discussed several key requirements for IoT devices to be able to utilize distributed identifiers and verifiable credentials, including (1) sufficient performance for cryptographic operations, (2) a sufficient amount of energy to perform the required operations, (3) non-volatile storage space to store the code and cryptographic keys, and (4) sufficient entropy source to generate random cryptographic keys. 
In~\cite{10.1145/3384943.3409436}, the authors presented \textsf{DIAM-IoT}, a decentralized identity and access management (IAM) framework for IoT that is capable of creating a unified, interoperable, and tamper-proof device identity registry on top of the blockchain by introducing DIDs and VCs into the lifecycle of IoT devices. In particular, the IEEE P2958 - Identity of Things Working Group~\cite{ieee_p2958} has started the standardization effort toward defining an IAM framework for IoT based on DIDs and VCs. 


Some recent studies have been proposed to address these challenges mainly from two aspects: delegation and optimization. For example, Lagutin et al. leveraged the ACE-OAuth (Authorization for Constrained Environments) framework to delegate the processing of VCs \cite{Lagutin2019}. In particular, the user who tries to access a resource-constrained IoT device (printer) presents his/her VC to the authorization server that handles the device's authorization request. The server verifies the VC-based authorization request and issues a proof-of-possession access token to the user to access the device. Nevertheless, the ACE framework does not cover how to exchange permissions and access tokens between the device and the authorization server. IoTChain addressed this by replacing the ACE-OAuth server with a smart contract deployed on a blockchain \cite{alphand2018iotchain}. In \cite{grande2020edge}, an edge-centric delegation of authorization was proposed to handle enrolment, access control, and roaming for resource-constrained IoT devices. In particular, the overhead was reduced by adopting lightweight crypto algorithms and outsourcing the authorization to edge devices. On the other hand, Fedrecheski et al. \cite{lopes2023low} proposed a set of techniques to optimize the use of DIDs in IoT environments \cite{lopes2023low}. Their DID scheme employs short identifiers and optimized cipher suites with smaller key sizes, aiming to decrease the size of DID documents. They utilized CBOR for document serialization, which reduces the document size up to four times compared to widely used JSON serialization.




Furthermore, various solutions have been proposed as the underneath verifiable data registry to support SSI. The authors of \cite{polychronaki2022identity} proposed a semi-decentralized blockchain-based IoT identity management framework to facilitate the creation of identities and the transfer of ownership of IoT devices. In addition, the proposed scheme also enables identity portability that can ensure secure device onboarding and transferring across networks. However, the adoption of Ethereum as the underlying blockchain can face severe performance challenges when serving a sheer amount of IoT devices. The direct interactions between IoT devices and the blockchain network also introduce non-trivial overhead to IoT devices, making it difficult to be adopted by resource-constrained IoT devices. To address such issues, recent studies have recognized IoTA tangle~\cite{luecking2020decentralized} and Hyperledger Indy~\cite{indy} as promising solutions particularly suitable for resource-constrained IoT scenarios. For example, the IoTA Tangle protocol, which is a distributed ledger based on directed acyclic graph (DAG) specifically designed to support resource-constrained IoT devices, has been recognized as a promising solution \cite{gebresilassie2020distributed}. The authors of \cite{luecking2020decentralized} proposed to leverage SSI and IoTA Tangle to manage the identities of IoT devices in a public and decentralized way. In addition, the ID management system is further enhanced via a Web of Trust approach, which enables automatic trust ratings of arbitrary identities. Furthermore, Hyperledger Indy, a decentralized ledger-based identity system, has been adopted in \cite{Lagutin2019} to implement an OAuth-based user authentication method, which moves the processing and management of DID and VC access policy to the authorization server to reduce overhead on resource-constrained IoT devices. Last but not least, to further reduce transaction costs to execute smart contracts on public blockchains, some studies proposed interconnecting multiple blockchains through interledger mechanisms so that smart contracts can be executed on private or permission blockchains at a lower cost. For example, the authors of \cite{siris2019interledger} introduced an authorization blockchain for smart contract execution and a payment blockchain to handle payments. However, the authorization functionality cannot be moved to the blockchain as it involves the processing of secret data. This further emphasizes the importance of a privacy-preserving authorization solution. 

\section{VC in IoT Context}
\label{sec:vciot}

\begin{figure}
    \centering
    \includegraphics[width=0.8\linewidth]{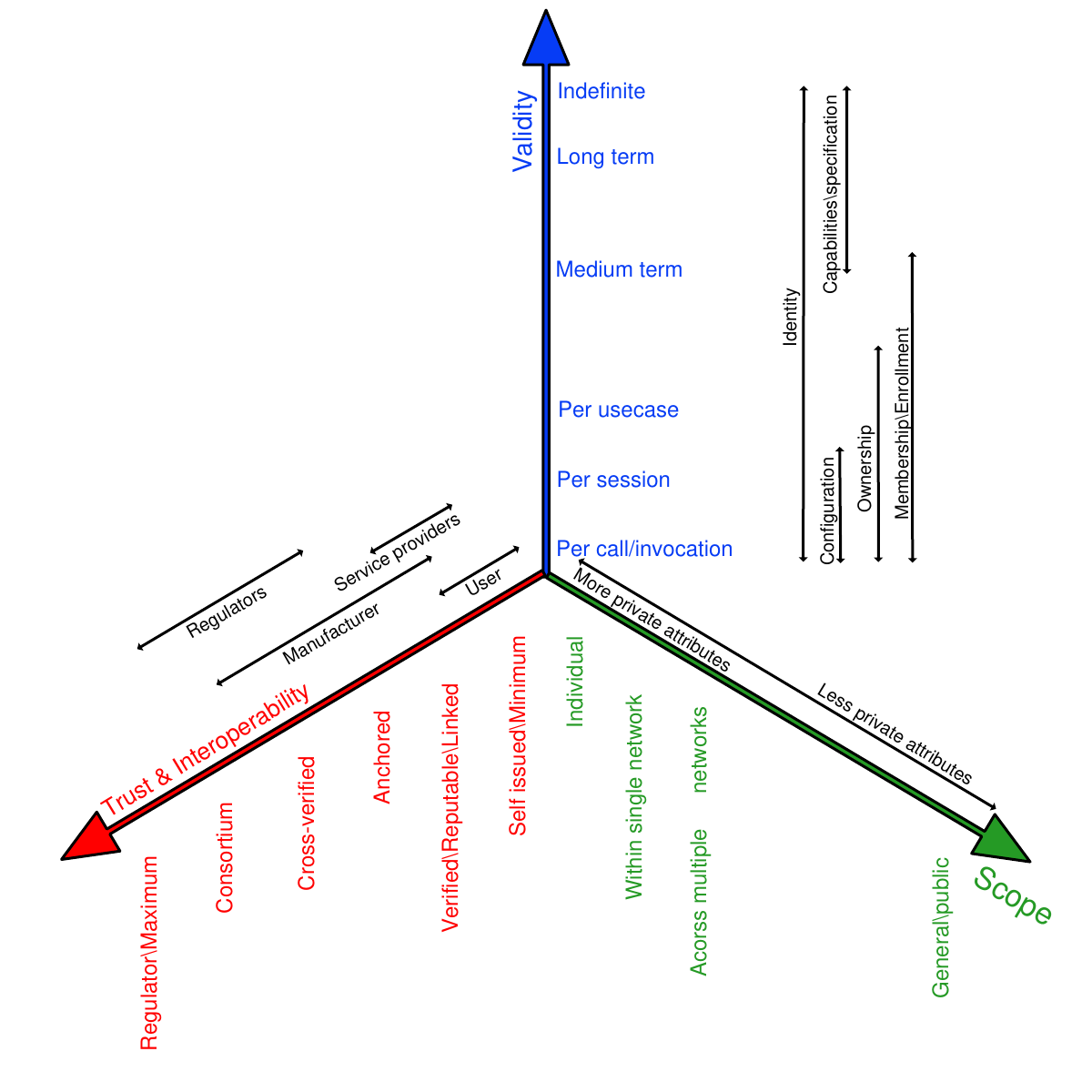}
    \caption{Proposed design matrix to aid VC formation in IoT Context}
    \label{fig:vc_iot_design}
\end{figure}

IoT devices serve a wide range of applications, each with unique requirements and constraints. A single VC per device cannot cover all the requirements and constraints of the device and its associated use cases. Therefore, IoT devices require multiple VCs that are specifically designed for specific purposes. Factors including the required level of trust, interoperability, acceptance, scope, the level of detail/attributes required, privacy, security, and validity must be considered when designing these VCs for IoT devices and their use cases to ensure the applicability, security, efficiency, and effectiveness of the VCs, use cases, and devices. This section introduces a design matrix that can be leveraged to design VCs in the IoT context. We categorized the essential factors into three primary categories: Trust \& Interoperability, Scope, and Validity as depicted in Figure \ref{fig:vc_iot_design}. 

Trust and interoperability of the VC is one of the essential factors covered by the design matrix, which also emphasise on the level of acceptance of the VCs. Categorizing VCs into different trust \& interoperable levels provides immense practical relevance to VCs in the IoT context as each use case requires a different level of trust and interoperability. A potential set of trust \& interoperable levels are depicted along in the red axis of the design matrix in Figure \ref{fig:vc_iot_design}. 

We consider "self-issued" to be the minimum level of trust \& interoperability required for credentials in the IoT context. This level typically originates from the device itself or from its owner/user. While this level offer some degree of autonomy, it is considered the least trustworthy due to their self-asserted nature. However, we denote that a much higher level of trust is possessed by the devices with trusted hardware modules such as Trusted Platform Module (TPM), Secure Element (SE), etc. A VC issued by a reputable entity or an entity that has already been verified is more trustworthy than self-issued VCs. These VCs benefit from external validation, adding a layer of trustworthiness to the claims of the IoT device. VCs anchored on a distributed ledger, such as a blockchain, provide a higher level of trust, enhancing its resistance to tampering and providing a trusted source for verifying the credential. A similar level of trust exists when a VC is linked to a robust, verifiable identity such as a public DID. Similarly to the PGP, an additional level of trust is embedded in VCs under the "cross-verified" level, as these VCs are independently endorsed by multiple reputable entities. However, it does not guarantee the higher level of interoperability since the level of interoperability is independent of the individual entities. "Consortium" highlights collective trust in particular aspects. These VCs are supported by a group of organizations or entities that work together, which signifies a high level of industry confidence in the credential and enables a higher level of interoperability. "Regulators" provide the highest level of trust and interoperability. VCs of this level are issued or endorsed by a regulatory authority, assuring the device's compliance and adherence to industry standards and regulations.

Similarly to the trust \& interoperability, scope must be carefully considered when formulating VCs for IoT devices. Some potential level of scope related to the IoT context is depicted along the green axis in Figure \ref{fig:vc_iot_design}. 

The scope of VC requirements can range from individual use cases to public scope. Identifying the correct scope enables issuers to determine the minimum set of attributes required in the credential to ensure privacy, whether it involves a specific representation of VCs allowing selective disclosures or zero-knowledge proofs. VCs with an individual scope should be unique, device-specific credentials tailored to the precise identity and capabilities of the IoT device. The individual scope ensures that each device can be distinguished and independently authenticated, which is crucial in cases where detailed, device-specific information and trust are of utmost importance. These credentials can contain a substantial number of claims about the device, making maximum privacy protection essential. Peer-wise DIDs/VCs and VCs for local access control are some of the applications of the individual scope where credentials are exchanged with another entity (device or user). 
For VCs used within the same network, the same level of individuality and privacy as the "Individual" scope may not be necessary. Nevertheless, they should still guarantee the necessary level of detail and identity within the specific network. The connected network scope encompasses VCs required when IoT devices need to communicate across multiple networks or even between different network domains. This scope should ensure compatibility across networks by identifying the minimum set of attributes to be included in the credential. These credentials should offer sufficient flexibility to accommodate various devices and networks while maintaining the required levels of privacy and security. Finally, the general/public scope represents the broadest scope, with minimal private information embedded in the credentials. These VCs are used mainly to demonstrate membership, compliance, or compatibility within the IoT context.

The concept of validity is another critical aspect that we have considered within our design matrix. In the context of VCs, validity refers to the duration of time during which a VC remains effective and the specific conditions governing its use. This aspect carries great significance in the IoT context due to the diverse range of potential applications. For instance, certain use cases necessitate VCs with relatively short lifespans, such as those intended for single-use scenarios or specific operations. These are particularly well-suited for situations requiring temporary access or authorization, like per-call/invocation or per-session VCs.
In contrast, more protracted use cases can benefit from medium-term VCs, which provide functionality throughout the duration of a particular session or use case. Long-term VCs offer an extended period of validation and are better suited for scenarios that require prolonged access or trust. Lastly, indefinite VCs represent the most enduring category, with no predetermined expiration date. They are typically used to denote immutable attributes of devices and use cases, including specifications, capabilities, and long-term identities. We also note that the same type of VC can be categorized under multiple validity periods, depending on the specific use of the credential. For example, the validity of an identity VC can span the entire spectrum, ranging from a single call to a long-term identity. The determination of the optimal validity period for a credential is a collaborative decision made between issuers and holders.  

\subsection{VC Issuers for IoT Devices}
Virtually any entity can act as a VC issuer according to the VC specification. However, the reliability, trustworthiness, and interoperability of a particular VC depend on the entity that issues it. This has led to the common practice of entrusting the issuance of VCs to well-established and trusted authorities. These authorities can take various forms, from government entities and educational institutions to manufacturers. However, the choice of VC issuer depends on the specific use case and the level of trust required for the credentials and its interoperability. In many cases, a combination of issuers, such as manufacturers for device-related attributes and government entities for regulatory compliance, can provide a comprehensive and trustworthy ecosystem of VCs. This section discusses potential VC issuers in the context of IoT devices and use cases.

\subsubsection{Manufacturers}
Manufacturers of IoT devices are among the most suitable options for VC issuers in the context of IoT. They have access to critical information about device specifications, capabilities, and identity, which is essential for creating accurate and reliable VCs that represent the device's identity and attributes. They further possess complete control over the device life-cycle, from production to distribution, enabling them to issue various types of VCs at each stage of the device's life-cycle. In addition, manufacturers are widely recognized as trusted entities with a reputation that they need to maintain to survive in the business. Consequently, VCs issued by manufacturers are typically trusted within the IoT ecosystem, as long as the manufacturer's reputation remains intact. Furthermore, generated VCs can be securely and efficiently transferred to the end users/owners, minimizing the risk of compromise by leveraging the manufacturers' supply chains. Additionally, existing legal frameworks and regulations mandate manufacturers to follow secure and efficient processes to adhere to industry standards. This ensures that the VCs they issue comply with the required industry standards.

In alignment with our proposed matrix, manufacturers have the flexibility to issue VCs that span a range of Trust \& Interoperable levels. These levels can vary from an individual manufacturer functioning at the "Verified" level independently to a more extensive "Consortium" level where a manufacturer is associated with a broader consortium. These VCs effectively span the entire spectrum of the scope axis, accommodating use cases that vary from individual scopes, such as manufacturer authorization, to general scopes that pertain to compliance or adherence to industry standards. Similarly, in the context of validity periods, manufacturers can issue VCs tailored to different validity and scope levels that encompasses addressing scenarios, which require shorter-term lifespans, like current configurations or session-specific VCs, to those necessitating extended duration such as long-term identity, specification, and capabilities. This versatility in VC issuance makes manufacturers an essential VC issuer in the IoT context.

\subsubsection{Regulators}
Regulatory entities have a unique advantage as VC issuers, given their authority to establish and enforce industry standards and regulations. This authority allows them to guarantee that VCs issued for IoT devices fully comply with the necessary industry standards, ensuring their reliability and security. Importantly, regulators are viewed as neutral and impartial entities with a primary objective of safeguarding users and consumers. This impartiality enhances the trustworthiness of the VCs issued by regulators compared to manufacturers who may have inherent biases.
In addition, regulators typically have a broader industry perspective and how it can evolve over time, which helps to issue well-designed and carefully considered VCs. Furthermore, this perspective enables high interoperability and uniformity in VC usage within the IoT ecosystem. Additionally, regulators have the ability to periodically assess compliance with issued VCs and ensure their integrity, given their monitoring and enforcement capabilities for compliance regulations.

Compared to manufacturers, regulators offer a higher level of trust and interoperability, stemming from their legal authority and oversight. In particular, regulators can issue VCs with Trust \& Interoperability levels ranging from "cross-verified" to the highest level achievable. The validity of VCs issued by regulators typically spans from medium to long-term, primarily due to the administrative overhead involved in issuing short-term VCs. These regulator-issued VCs often encompass a broader scope, including those that traverse multiple networks and fall under the general scope category.

\subsubsection{Service Providers}
Service providers can be considered another potential VC issuer in the IoT context. However, it's important to note that service providers can only issue more granular and specific VCs relate to their service, while those from manufacturers and regulators tend to be broader and more generalized. The level of interoperability and trustworthiness associated with the VC depends on the reputation and the quality of service provided by a particular service provider. 

In terms of mapping to the matrix, service providers generally offer a more limited range of Trust \& Interoperability levels compared to manufacturers and regulators, typically spanning from "verified" to "anchored". These VCs are typically intended for use within the service provider's network and a select few connected networks that share an affiliation with the service provider, similar to the approach used by platforms like Google or Facebook for third-party logins. In terms of validity, these VCs have a shorter lifespan when compared to VCs issued by more highly trusted issuers.

\subsubsection{Owners}
Device owners can facilitate the required VCs given the possibility that IoT devices are enrolled in different processes and used by multiple users. Owner-issued VCs help ease the implementation of appropriate access control mechanisms to the services that they provide without relying on third party providers.

The trust and interoperability levels provided by device owners are typically minimum, falling into the categories of self-issued or linked when their identity is well established. In either case, interoperability remains at a basic level. Additionally, these VCs often have limited scope, primarily for individual use cases or within the same network controlled by the owner. However, owners have the flexibility to issue VCs across the entire validity spectrum.

\subsection{Common VC for IoT Use Cases}
There are numerous use cases for IoT devices in various domains. Nevertheless, certain common fundamental functionalities can be identified among most of these use cases. In this section, we categorize these common functionalities in IoT use cases as follows and establish VCs for each of the identified functionalities. This categorization and structuring of VCs are intended to facilitate future standardization efforts and prevent the need for multiple schemes for the same functionality. We also align these VCs within the framework of our proposed design matrix. 

\begin{center}
\begin{forest}
  for tree={
    align=center,
    font=\sffamily,
    edge+={thick, -{Stealth[]}},
    l sep'+=10pt,
    fork sep'=10pt,
  },
  forked edges,
  if level=0{
    inner xsep=0pt,
    tikz={\draw [thick] (.children first) -- (.children last);}
  }{},
  [Common Functionalities in IoT Context
    [Identity]
    [Ownership]
    [Communication]
    [Capabilities]
    [Configurations]
    [Onboarding]
  ]
\end{forest}
\end{center}

\subsubsection{IoT Identity}
The identity of an IoT device is an integral part of any IoT use case. In particular, identity stands as a paramount factor in the context of security, as it helps establish trust in the device and its data. A secure and reliable identity helps enforce effective access controls and appropriate monitoring, mitigating numerous security concerns. Furthermore, a flawless identity entails capturing unforgivable attributes or properties of the IoT device that firmly bind the identity to the device, such as a Physical Unclonable Function (PUF). Despite the advantages of a flawless identity, universal adoption of such an identity is impractical due to factors such as the heterogeneous nature of IoT devices and associated costs. Therefore, alternative approaches are required to establish robust identities for IoT devices without being limited to physical attributes.

We argue that IoT devices should employ multiple identities to cater to various use cases with varying validity periods, from short-term to indefinite. Similarly, the trustworthiness and interoperability of the identity can span from a self-issued identity for peer-to-peer interactions to the maximum trust and interoperable level. In addition, identity is essential in any context of scope to establish proper identification. To establish the device's static identity, we have leveraged commonly available attributes of IoT devices to define the VC of the static/indefinite/manufacturer IoT identity, as illustrated in Listing \ref{lst:vc_identity}. This VC essentially encapsulates the device's DID along with a collection of static attributes, including device characteristics and manufacturer details. These attributes remain consistent throughout the device's operational lifespan. We excluded the device's public key as a separate attribute because it can change throughout the device's life cycle. The current public key of the device can be obtained from the device's DID document, which is linked to the device's DID.

\begin{lstlisting}[caption = Sample VC for a static IoT identity,captionpos=b,label=lst:vc_identity,language=json]
{
  "@context":[..],
  "id": "did:iot:device:vc:123456789",
  "type": ["VerifiableCredential", "StaticIoTIdentityVC"],
  "issuer": "did:iot:manufacturer:123456789",
  "validFrom": "2023-04-01T10:11:12Z",
  "credentialSubject": {
    "id": "did:iot:device:123456789",
    "serialNo": "123456789",
    "manufacturedDate":"2022-12-01T00:01:02Z",
    "manufacturer": "did:iot:manufacturer:123456789",
    "modelNo": "XYZ",
    "batchNo": "12345"
  },
  "credentialSchema": {..},
  ...
  "proof": {..}
}
\end{lstlisting}

The static/indefinite/manufacturer identity VC must be issued only by the corresponding device manufacturer. The initial identity VC for an IoT device is issued and digitally signed by the device manufacturer before the device leaves the factory. Possession of the manufacturer's digital signature thwarts adversaries' attempts to mimic fake or non-existent IoT devices. Depending on the device's storage capacity, this identity VC can be either embedded within the device or stored in a secure location. If the device lacks the capacity to store its VCs, the VC must be securely transferred to the end user through the device supply-chain.

A separate short-to-long-term identity VC can be constructed by integrating specific device metadata, such as the current firmware version and the timestamp of the last update, into the `Static Identity VC' to provide additional insights into device status, as illustrated in Listing \ref{lst:vc_dyn_identity}. The validity of this VC depends on the frequency of firmware updates. If the device's currently running firmware version changes, a new VC must be issued. We assume that the manufacturer has a secure and reliable method, such as remote attestation, to verify the current firmware and update date/time of a device without relying on user-provided information. When a new VC is issued, the existing identity VC should be revoked. This process ensures that the device identity information remains accurate and up-to-date. Additionally, the history of firmware updates can be appended to this VC to provide a more reliable assessment of device trustworthiness, considering that a device could be compromised if the owner delays applying the latest security update.

\begin{lstlisting}[caption = Sample VC for an IoT identity with dynamic information,captionpos=b,label=lst:vc_dyn_identity,language=json]
{
  ..
  "credentialSubject": {
    "id": "did:iot:device:123456789",
    "firmwareVersion": "1.2.3",
    "lastUpdatedDate": "2023-04-01T00:01:02Z"
  },
  ..
}
\end{lstlisting}

Additional dynamic attributes such as geolocation, operating system can be embedded to the VC depending on the form of the device (wearable, edge, embedded, etc.). However, we note that not all IoT devices possess such attributes.

\subsubsection{IoT Ownership}
Many IoT use cases involve representing the ownership of an IoT device, which associates the device with its owner. An accurate representation of the ownership of the device can prevent adversaries from misuse of the device. Additionally, changes in device settings and firmware updates can be managed through appropriate access control mechanisms, given the availability of accurate ownership details of the IoT device. 

The proposed ownership VC is shown in Listing \ref{lst:vc_ownership}. It includes the device DID, the owner DID, and the purchase date. Additionally, transaction details can be embedded in the VC if required, given that the transaction is recorded on a distributed ledger. Despite the additional workload imposed on manufacturers, they are the most suitable VC issuers for the ownership VC. Manufacturers can implement a smart contract or a standard API to streamline the ownership transfer process. Using digital signatures from both the buyer and the seller with the existing ownership VC can help reduce manual interventions and ensure the authenticity of the transaction. It also allow revocation of the existing ownership VC as the issuer remain consistent throughout the device life cycle.
Similarly to the identity VC, the initial ownership VC can be issued by the manufacturer to either itself or authorized sellers. Subsequently, owners/sellers must use the deployed API to provide a new VC reflecting the new ownership and revoking existing ownership when selling the device.

\begin{lstlisting}[caption = Sample VC for an IoT ownership,captionpos=b,label=lst:vc_ownership,language=json]
{
  "@context":[..],
  "id": "did:iot:device:vc:own:123456789",
  "type": ["VerifiableCredential", "IoTOwnershipVC"],
  "issuer": "did:iot:manufacturer:123456789",
  "credentialSubject": {
    "deviceId": "did:iot:device:123456789",
    "owner": "did:iot:user:123456789",
    "purchasedDate" : "2022-01-01"
  },
  ..
}
\end{lstlisting}

Unlike the static identity of a device, ownership can change throughout the device's lifespan. Consequently, the validity of ownership typically spans from short to medium term. However, there are situations where ownership can be long-term or even indefinite, particularly in cases where the IoT device is installed in remote fields or mission-critical environments. The trust level of ownership should not be self-issued, as self-issued ownership can pose various issues, including device theft. The scope of ownership should be relevant, ranging from an individual to a general scope. However, the privacy aspect of ownership should be a concern depending on the scope.

\subsubsection{IoT Communication}
IoT communication methods encompass a broad spectrum of technologies and protocols, each designed to meet specific needs in various IoT applications. These methods enable data exchange among interconnected devices, sensors, stakeholders, or centralized systems, making them a fundamental component of interconnected networks such as smart homes, industrial IoT, and smart cities. We propose defining a communication VC for IoT devices by categorizing communication into four primary groups: wired, wireless, cellular, and satellite, as shown in Listing \ref{lst:vc_comunication}.

\begin{lstlisting}[caption = Sample VC for an IoT comunication,captionpos=b,label=lst:vc_comunication,language=json]
{
  "@context":[..],
  "id": "did:iot:device:vc:comm:123456789",
  "type": ["VerifiableCredential", "IoTComunicationVC"],
  "issuer": "did:iot:manufacturer:123456789",
  "credentialSubject": {
    "deviceId": "did:iot:device:123456789",
    "wired": {
        ethernet
    },
    "wireless": {
        bluetooth,
        Zigbee,
    },
    "cellular": {
        5G
    },
    "satellite": { ..}
  },
  ..
}
\end{lstlisting}

Wired communication section lists all physical connections, such as Ethernet, that are supported by the device. Most IoT devices use wireless technologies, including Wi-Fi, Bluetooth, Zigbee, and Z-wave. Some IoT devices require cellular communication to provide extensive coverage and reliable connectivity. Additionally, there are IoT devices that are deployed in isolated or remote areas that require satellite communication. Moreover, the specific mode or version of the communication protocol such as Bluetooth 4.0, 5G-SA can also be included in the credential.

The manufacturer can generate the Communication VC, specifying the different protocols that an IoT device supports, before the device leaves the factory and securely stores it. This VC is assumed to be immutable to the device, unless the device possesses pluggable interfaces to support additional protocols or malfunctions of the supporting protocols. Thus, the validity of this credential is intended to be long-term to indefinite. It is ideal to have a higher level of trust and interoperability for the Communication VC. However, the level of trust and interoperability depend on the manufacturer. IoT device manufacturers can collectively form a consortium or enroll in an existing consortium to agree upon a well-established and recognizable layout for the VC, increasing interoperability. Entities seeking to communicate with the IoT device can reference its Communication VC to determine the most appropriate protocol for their requirements. Hence, the scope extends beyond the individual scope.

The overhead of issuing separate communication VCs for each individual device can be alleviated by issuing a common VC for a specific batch or model of devices, given that manufacturers often produce large quantities of devices in a single batch, and devices in the same batch or model usually share the same communication capabilities. This common VC can then be stored in a publicly accessible repository, such as a public blockchain, instead of being associated with each device individually. Stakeholders can retrieve the corresponding communication VC for a specific device from the public repository by leveraging the device identity provided in the identity VC. However, a common communication VC does not reflect the true status when a new communication module is installed on the device or an existing one malfunctions.

\subsubsection{IoT Capabilities}
Similarly to communication, IoT devices exhibit a diverse range of capabilities, each of which contributes to their functionality in various contexts. These capabilities encompass their ability to process information, store data, and perform other specialized functions. We propose categorizing these capabilities into three distinct categories: computation, memory, and other, as depicted in Listing \ref{lst:vc_capabilities}. 

\textbf{Computation Capabilities}: The computation power of the IoT device allows processing of collected or received data and making decisions based on the information they collect. These capabilities can vary widely depending on the device. Potential attributes include information about the processor, clock speed, power consumption, etc.

\textbf{Memory Capabilities}: The memory capabilities of the IoT devices encompass their ability to store data, code, and configurations. This includes both volatile and non-volatile memory, such as RAM and flash storage. Potential attributes include the available type of memory, memory size, bandwidth, etc.

\textbf{Other Capabilities}: This category provides the flexibility to encompass additional capabilities that may not fit neatly into the other three categories. These capabilities can vary widely depending on the specific type and purpose of the IoT device. Examples might include sensor capabilities (e.g., temperature, humidity, motion), power management features, security protocols, and support for various industry-specific standards. 

We anticipate that manufacturers will issue capability VCs before the IoT device leaves the manufacturing facility. Similarly to communication VCs, this VC can be stored on the IoT device, shared with the user through the supply chain, or a common VC can be issued for a batch or model of devices and stored on a public repository. A common capability VC can raise the same concerns as a common communication VC, given the possibility of hardware upgrades and malfunctions. Capability VCs provide a standardized way to document and share the capabilities of IoT devices, ensuring that the device can be effectively integrated into various ecosystems.

\begin{lstlisting}[caption = Sample VC for an IoT capabilities/specification,captionpos=b,label=lst:vc_capabilities,language=json]
{
  "@context":[..],
  "id": "did:iot:device:vc:cap:123456789",
  "type": ["VerifiableCredential", "IoTCapabilityVC"],
  "issuer": "did:iot:manufacturer:123456789",
  "credentialSubject": {
    "deviceId": "did:iot:device:123456789",
    "computation": {
        clockSpeed: <cpu clock speed>,
        cpuArch: <cpu architecture>,
        noOfCores: <number of cores>,
        ..
    },
    "memory": { ..},
    "other": { ..}
  },
  ..
}
\end{lstlisting}

The validity of this credential is intended to be long-term to indefinite given the fact that the device capabilities remains unchanged in many scenarios. Manufacturer-provided trust is sufficient in many scenarios for the `capability VC'. Nevertheless, some attributes, such as compliance to a particular industry standard, may require additional level of trust from issuers such as standard organization or a consortium. Similarly to communication VC, the interoperability of the `capability VC' can be improved by association with other manufacturers and consortiums. Similarly, the scope extends beyond the individual scope. 

\subsubsection{IoT Configurations}
While many IoT devices lack configurable settings, some do possess configurations, necessitating the definition of a standardized approach to express these configurations. Given the diverse range of configurations and use cases, it is impractical to define a common set of attributes. Despite this lack of uniformity, device manufacturers can define their set of attributes that are deemed appropriate to share and attest to together with the device identity. Thus, we propose to group the available configurations into five categories: thresholds, security, communication, user, and other as shown in Listing \ref{lst:vc_config}.

The threshold category should define all the configuration parameters that serve as limits of the device such as maximum throughput, minimum latency, etc. All the security-related configurations, such as particular encryption algorithms, digital signatures, and access control mechanisms, must be defined inside the security category. Communication settings can include port numbers and protocols. Settings such as device name, user-specific power schedules, and other user-defined configurations can be defined inside the user category. Any other available settings must be defined under the other category.

\begin{lstlisting}[caption = Sample VC for an IoT configuration,captionpos=b,label=lst:vc_config,language=json]
{
  "@context":[..],
  "id": "did:iot:device:vc:config:123456789",
  "type": ["VerifiableCredential", "IoTConfigVC"],
  "issuer": "did:iot:manufacturer:123456789",
  "validFrom": "2023-04-01T10:11:12Z",
  "credentialSubject": {
    "deviceId": "did:iot:device:123456789",
    "thresholds": { ..},
    "security": { ..},
    "communication": { ..},
    "user": { ..},
    "other": { ..}
  },
  ..
}
\end{lstlisting}

The attestability of selected attributes is a pivotal criterion for the `Configuration VC', as users can freely alter the default configuration. Thus, the device should support remote attestation (software, hybrid, or hardware) to be eligible for possessing a `Configuration VC'. Alternatively, IoT devices with trusted hardware modules can generate self-issued Configuration VCs.

Manufacturers are optimal issuers for `Configuration VCs', as they possess remote attestation capabilities for the device unless the device possess trusted hardware modules. Thus, the trust level of the VC can range from Verified to Consortium, depending on the manufacturer. Unlike other VCs, the validity of this credential is on the shorter side, being valid only for a single invocation or per session. The scope is also limited to a single network or over several connected networks.

\subsubsection{IoT Onboarding}
Device onboarding is an integral part of many IoT use cases. The `onboarding VC' should contain all the required information and configuration, such as the device identity, manufacturer information, initial configuration settings, supported onboarding protocol, and onboarding service information, to ensure a secure and efficient onboarding process that minimizes human intervention~\cite{fagan2023trusted,cooper2021fido}. We have divided the `onboarding VC' attributes into two high-level groups: onboardee and onboarder. Given the availability of diverse onboarding protocols, we propose grouping the device-specific attributes, including device identity, configurations, ownership, and other required information, under the onboardee category. Mutual authentication is required in many onboarding scenarios. Thus, we are proposing to define the onboarding service attributes, including its identity and other relevant information, in the onboarder section.

\begin{lstlisting}[caption = Sample VC for an IoT onboarding,captionpos=b,label=lst:vc_onboard,language=json]
{
  "@context":[..],
  "id": "did:iot:device:vc:onboard:123456789",
  "type": ["VerifiableCredential", "IoTOnboardingVC"],
  "issuer": "did:iot:manufacturer:123456789",
  "credentialSubject": {
    "deviceId": "did:iot:device:123456789",
    "onboardee": {
        "identity": { ..},
        "configuration": { ..},
        "ownership": { ..},
        "other": { ..}
    }
    "onboarder": { 
        "identity": { ..},
        "other": { ..}
    }
  },
  ..
}
\end{lstlisting}

Manufacturers or relevant service providers can issue the `onboarding VC' for IoT devices, providing a higher level of trust and interoperability than self-issued credentials. The `onboarding VC' should be designed with a relatively short validity period to ensure that the device's information is current and minimize the risk of misuse. The scope of the credential should be limited to the given network. Possession of a comprehensive `onboarding VC' allows IoT devices to seamlessly integrate into diverse environments while maintaining a high level of security and interoperability.

\section{VC Life-cycle Management in IoT Context}
\label{sec:lifecycle}

\begin{figure}[!htp]
    \centering
    \includegraphics{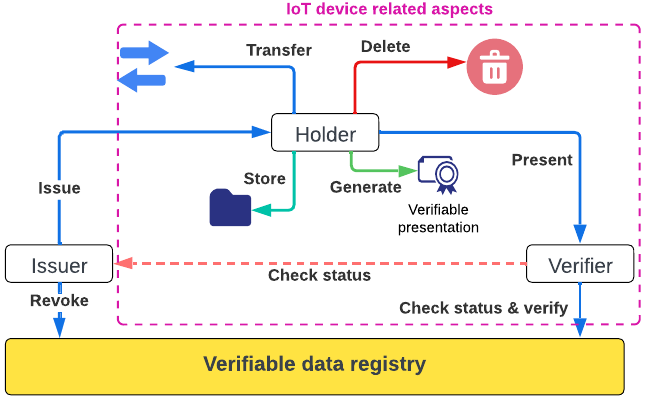}
    \caption{Verifiable credential life-cycle and its IoT aspect}
    \label{fig:vc_lifecycle}
\end{figure}

The life-cycle of VC management involves several processes, including issuing, presenting, and verifying, as illustrated in Figure \ref{fig:vc_lifecycle}. In the context of IoT devices, we focus on the processes that require interactions with these devices as the holder and verifier, including storing, transferring, presenting, generating verifiable presentations, status checking, and verifying (enclosed within the dashed line). The remaining actions (issuance and revocation) are presumed to be handled by trusted entities such as manufacturers or regulatory bodies. 
A single approach to manage the VC life-cycle (related aspects) is impractical in the IoT context due to the diverse capabilities of IoT devices. Thus, multiple capability-aware methods are required to effectively manage VC for IoT devices. This section describes some of the identified potential methods.

\subsection{Smart Autonomous IoT Devices}
The concept of autonomous IoT devices independently managing VCs, much like humans, is plausible in the near future with the growing sophistication of smart IoT technology. To achieve this capability, IoT devices need embedded secure storage for VC storage, robust communication capabilities to interact with issuers and verifiers, and advanced, optimized cryptographic libraries to independently handle generating Verifiable Presentations (VPs) and verifying VCs without relying on external servers or centralized authorities. 
These autonomous IoT devices must also possess the ability to assess trust based on the information contained within the VCs. This may involve tasks such as verifying the authenticity of firmware updates or distinguishing between their owner and other users within a network.

The emergence of autonomous IoT devices with VC management capabilities will enable several significant advantages. These include reducing latency, and enhancing privacy \& security throughout the VC management process. These advantages make autonomous IoT devices well-suited for applications where real-time, independent trust evaluation is critical to operational success. Nevertheless, factors such as power utilization associated with connectivity (to distributed ledgers) and other computations, such as verification, need to be considered to ensure better power management of the device.

\subsection{Resource Constrained IoT Devices}
Despite the plausibility of full autonomous IoT devices that are capable of managing the entire VC life-cycle, most IoT devices rely on their owner/user to manage VCs. This delegation rely on factors such as the deployment environment and device capabilities. Potential delegation methods include smart contracts, cloud solutions, edge devices, proxies, and existing frameworks such as ACE-OAuth~\cite{Lagutin2019} and OpenID~\cite{chadwick2022openid}. 

Leveraging smart contracts is a robust and transparent method to manage VCs in the resource constrained IoT context given the availability of light weight APIs to interact with smart contracts. Smart contracts can serve as repositories for VCs, ensuring their tamper-proof storage on the blockchain for IoT devices that do not possess sufficient storage capabilities. VCs stored in a blockchain are immutable and preserve integrity. However, not every VC can be stored on blockchain as it can jeopardize the privacy given VCs can contain private information associated with an IoT device. Smart contracts can enable the seamless transferring of VCs between different entities in situations such as when the ownership need to be updated. IoT devices can further leverage smart contracts to share VCs, and to verify the current status and authenticity of VCs. Additionally, smart contracts can also generate VPs enclosing VCs on behalf of IoT device given the possibility of smart contract-based signatures such as ERC-1271 \cite{erc1271}.

Use of edge devices, proxies or cloud solution including IoT gateways to tackle the VC life-cycle is another possibility. These solutions can serve as secure storage repositories for VCs. They can maintain a local cache of VCs on behalf of IoT devices, ensuring that the credentials are readily available. The edge device can be synchronized with the issuing authority or blockchain periodically to maintain the integrity and validity of the credentials. 
VC transfer, including validation and secure transmission, can be handled by associated edge devices to minimize the energy and bandwidth consumption of IoT devices. Further, status and validity checks can be performed by the edge device initiating the communication with entities, including the blockchain or trusted authorities. Additionally, it is more practical to verify VCs on behalf of IoT devices during interactions with verifiers. When a verifier requests proof of a credential, the IoT device can delegate this task to the edge device. The edge device can then perform the verification, ensuring that the VC is valid and correctly presented. In the context of generating VPs, edge devices can perform the cryptographic operations required to create VPs more conveniently than smart contracts, allowing resource-constrained devices to simply request a VP from the edge device when authentication or verification is required. 

Delegation of VC management to IoT device owners is another alternative for resource-constrained IoT devices. Device owners can maintain a centralized repository or digital wallet for VCs associated with their IoT devices. These wallets or repositories can be hosted locally on a PC, laptop, mobile phone, or in a cloud storage, ensuring the required levels of accessibility and security. The device owner can act as an intermediary for transferring VCs of their IoT devices, where the device only needs to communicate with its owner, reducing the attack surface and enhancing trust. Moreover, it also minimizes network connectivity for the IoT device, which helps to reduce power usage. Owners (or their wallets) can be programmed to pull the latest status and updated VCs from the issuers or blockchain to maintain the validity of device VCs. Similarly, the owner or their associated computing devices can act as the verifier of the IoT device and conduct VC verification by interacting with the blockchain or other authorized entities, reducing the communication requirements for the devices. Furthermore, the device owners can generate VPs for the IoT VCs, as they possess the necessary capabilities for required cryptographic operations, including signing and proof generation. Moreover, the owner's key can be leveraged in generating VPs instead of directly accessing device keys.

Additionally, delegation of VC management to edge devices, gateways, cloud solutions and smart contracts increases the security of VCs as many IoT devices can be easily compromised compared to other standard devices unless the device possess trusted hardware modules. 

\section{Optimizing SSI for IoT Context}
\label{sec:optimization}

\subsection{Encoding \& Serialization}
Data exchange formats are an essential factor in realizing IoT-friendly SSI in the IoT context that ensures efficient communication and interoperability between diverse devices and systems. In particular, some of the commonly used IoT communication protocols such as BLE and LoRA have less than a few hundred bytes~\cite{lopes2023low}. Thus, choosing efficient encoding and serialization formats can reduce the communication overhead and minimize the storage and processing requirements. 

In recent years, Concise Binary Object Representation (CBOR) \cite{cbor_spec} has emerged as a better choice to represent VCs~\cite{lopes2023low,young2021verifiable,fotiou2022iot} and DID documents~\cite{cbor_diddoc}. Unlike traditional JSON, which uses UTF-8 encoding (by default) to ensure human-readable text, CBOR is a binary-encoded format. This binary encoding results in more compact data representations, reducing the size of VCs and DID documents transmitted over IoT networks. This efficiency is essential in many IoT scenarios where bandwidth and storage resources can be limited. Additionally, CBOR provides faster serialization and deserialization processes, which is beneficial for resource-constrained IoT devices with limited processing capabilities.

Similarly to CBOR, binary JSON (BSON)~\cite{bson_spec} also offers significant advantages over JSON in representing VC via binary encoding in the IoT context. Although CBOR is designed specifically for IoT devices, BSON was originally designed for databases, specifically for MongoDB. Furthermore, compactness was not a primary objective in the BSON design. Therefore, BSON-encoded VCs can occupy more space than JSON-encoded ones, depending on the content of the VC or DID document. However, since its encoding approach is closer to machine representations, it is more efficient than JSON in terms of parsing, which reduces processing overhead for constrained devices~\cite{kortesniemi2019improving}.

\subsection{Communication}
Communication protocols are another critical aspect in implementing IoT-friendly SSI. Several factors must be considered when selecting the optimal communication protocol, including security, efficiency, scalability, and interoperability.

Constrained Application Protocol (CoAP)~\cite{coap_spec} is a protocol specifically designed for resource-constrained devices, making it highly efficient in terms of both bandwidth and power consumption. CoAP supports the RESTful architecture, enabling seamless integration with web services and Internet-based services. This RESTful approach facilitates the exchange of VCs and DID documents with relevant entities, including holders, verifiers, and issuers, similar to a standard web-based API approach. Additionally, CoAP leverages User Datagram Protocol (UDP), reducing message size and handshaking compared to protocols that use Transmission Control Protocol (TCP)~\cite{mahalle2021oauth}. This asynchronous communication provides essential support for managing VCs in the IoT context, where device availability and connectivity are often limited or restricted. Furthermore, built-in security modes can be utilized to provide end-to-end security when VCs and DID documents are exchanged, eliminating the need for additional layers of security. These advantages collectively make CoAP a robust choice for SSI communication in the IoT, enabling efficient, lightweight, and secure interactions in resource-constrained environments.

Message Queuing Telemetry Transport (MQTT~\cite{mqtt}), another protocol commonly used in IoT communication, is specifically designed for IoT use cases. It requires low energy and bandwidth and allows for real-time response. Unlike CoAP, MQTT is based on the TCP/IP protocol. However, there is a variant, MQTT-SN, that uses other protocols such as UDP or Bluetooth \cite{abdelrazig2021blockchain}. The use of publisher-subscriber architecture along with TCP makes MQTT more reliable and scalable compared to CoAP. Additionally, unlike CoAP, MQTT is an asynchronous protocol, minimizing connectivity constraints on IoT devices where devices need to stay online. In addition, MQTT's message overhead is two bytes, whereas in CoAP it is four bytes. These advantages make MQTT a suitable choice for SSI communication in the IoT context.

DIDComm~\cite{didcomm} is a protocol specifically designed for use cases involving SSI. It utilizes DIDs to establish communication between entities. Unlike MQTT and CoAP, DIDComm is independent of transport protocol and allows authenticated message exchanges and message routing over loosely trusted routers~\cite{lopes2023low}. However, DIDComm uses JSON serialization, which has a slight overhead compared to protocols that use binary serialization. Moreover, some recent studies have shown that DIDComm exhibits higher latency, although resource efficiency is comparable~\cite{sharma2022blockchain}. Nevertheless, DIDComm is suitable in IoT use cases where higher level of security is required as it provides the end-to-end security, which MQTT and CoAP lacks.  

\section{Discussion}
\label{sec:discussion}

The concept of SSI is gaining traction with the standardization of DIDs, VCs, and the emergence of real-world SSI implementations such as the European Union's eIDAS~\cite{schwalm2022eidas} and British Columbia's "Tell Us Once"~\cite{pohn2021eid} projects. Further, the recent shift towards the zero-trust~\cite{zerotrust} architecture is beginning to incursion into the IoT context, where implicit trust is absent, unlike other trust models. Stakeholders need authentication to access diverse resources with varying privileges over different time intervals. This presents new challenges, such as the need for more frequent credentials, shorter validity periods, and more attributes. IoT-friendly SSI solutions can significantly help in implementing zero-trust models in the IoT context. 

Scalability is a key concern for IoT-friendly SSI, given the vast number of devices. Many SSI implementations rely on blockchains, but traditional consensus mechanisms on popular blockchains are not effective for large-scale IoT use cases due to low throughput and high latency. Therefore, consensus mechanisms that allow higher throughput rates and low latency, blockchains specifically designed for SSI purposes, such as Hyperledger Indy~\cite{indy}, or alternative distributed ledger mechanisms such as IOTA Tangle~\cite{luecking2020decentralized}, should be leveraged in designing more scalable SSI solutions for the IoT context.

Similarly to SSI in the human context, the use of the same DID/VC can raise privacy concerns for IoT devices and their owners/users. Privacy-preserving techniques, such as anonymization and pseudonymization, and zero-knowledge proofs are vital for certain IoT use cases. Ensuring that these methods are IoT-friendly is essential, considering the constraints in the IoT context.

Despite the advantages of SSI, the dependence on Internet/distributed ledger connectivity can hinder IoT-friendly SSI for devices with limited Internet access. Alternative approaches such as continuous attestation of VCs~\cite{offlineverify} are necessary to ensure accurate verification, confirming that the presented VCs / DID documents are up-to-date and not revoked. Additionally, intelligent caching approaches can be implemented to cache DID documents locally to limit network access to ensure that devices do not connect the distributed ledger unnecessarily. 

Moreover, unlike human-centric approaches, transitioning existing identity solutions to SSI-based approaches in the IoT context is not straightforward due to the diverse use cases and characteristics of existing IoT devices. Therefore, a comprehensive infrastructure, encompassing standardization, regulations, and more accessible open-source implementations, must be established to facilitate and incentivize the adoption of SSI in current IoT use cases.

\section{Conclusion \& Future Work}
\label{sec:conclusion}

Although numerous initiatives apply the novel concept of SSI in human-centric contexts, the diversity of devices, resource limitations, and other constraints in the IoT realm present challenges to the widespread adoption of IoT-friendly SSI. Addressing these limitations is essential for the effective integration of SSI into the broader IoT context. This work proposed a design matrix for designing VCs in the IoT context and analyzed effective ways of adopting the SSI concept in the IoT domain, including generalizing IoT VCs, life-cycle management, and optimizations for IoT-friendly SSI.

Future work needs to explore additional approaches to adopt SSI in more resource-constrained devices, especially in scenarios involving IoT devices with limited or no network connectivity. Additionally, there is a need for more efficient encoding, serialization, and communication methods to minimize resource utilization in the IoT domain.

\bibliographystyle{ACM-Reference-Format}
\bibliography{references}

\end{document}